\documentclass[a4paper,11pt,english]{article}
\usepackage{amsfonts}
\usepackage{graphicx}
\usepackage{amsmath}
\usepackage{color}

\newcommand{\be}{\begin{equation}}
\newcommand{\ee}{\end{equation}}
\begin{document}


\begin{titlepage}
\begin{center}

\noindent{{\LARGE{Colored black holes and Kac-Moody algebra}}}

\smallskip
\smallskip

\smallskip
\smallskip

\smallskip
\smallskip
\smallskip
\smallskip
\smallskip
\smallskip
\smallskip
\smallskip

\noindent{\large{Gaston Giribet, Luciano Montecchio}}

\smallskip
\smallskip

\smallskip
\smallskip

\smallskip
\smallskip

\centerline{Physics Department, University of Buenos Aires FCEyN-UBA and IFIBA-CONICET}
\centerline{{\it Ciudad Universitaria, pabell\'on 1, 1428, Buenos Aires, Argentina.}}

\end{center}

\bigskip

\bigskip

\bigskip

\bigskip

\begin{abstract}
We demonstrate that the near horizon symmetries of black holes in Einstein--Yang-Mills (EYM) theory are generated by an infinite-dimensional algebra that contains, in addition to supertranslations and superrotations, a non-Abelian loop algebra. This means that the Virasoro--Kac-Moody structure of EYM in asymptotically flat spacetimes has an exact analog in the near horizon region. 
\end{abstract}
\end{titlepage}


\section{Introduction}

In 2015, Hawking conjectured that, in their near horizon limit, black holes might exhibit an infinite-dimensional symmetry \cite{Hawking} similar to the supertranslations that appear near null-infinity in asymptotically flat spacetimes \cite{BMS, BMS2, BMS3}. His conjecture was motivated by the recent developments connecting the subjects of asymptotic symmetries, soft theorems and memory effect in gravity and gauge theories \cite{Strominger}. The original idea was that horizon supertranslations could have something to do with the black hole information puzzle \cite{HPS}, an interesting idea that gave rise to a thoughtful debate \cite{Porrati1, Porrati2, Porrati3}. Nevertheless, regardless whether relevant or not for the information loss problem specifically, the emergence of infinite-dimensional symmetries near the black hole event horizons turned out to be an interesting discovery on its own right, as it suggests that there might still be important lessons to be learnt from symmetries about the infrared structure of gravity and gauge theories.

The existence of infinite-dimensional symmetries in the vicinity of black hole horizons was made precise in \cite{DGGP}, where it was shown that, in addition to supertranslations, black holes also exhibit superrotations in their cercanity. Further details of these symmetries and their associated charges were given in \cite{DGGP2, HPS2} and in references thereof. The study of infinite-dimensional symmetries in the near horizon region has antecedents \cite{Koga, Hotta, Hotta2}, and more recently it led to interesting developments and generalizations; see for instance \cite{DGGP3, DonnayMarteau, Grumiller, otro, Penna, Pu, Wald, oder1, oder2, oder3}.  

It was shown in \cite{DGGP3} that the addition of Abelian gauge fields results in a further enhancement of the near horizon symmetry, yielding a new set of supertranslation currents for each $U(1)$ commuting factor in the gauge group, at least. Here, we will consider the case of Einstein gravity coupled to Yang-Mills theory for an arbitrary gauge group $G$, and we will show that, in addition to supertranslations and superrotations, the black holes of the theory exhibit an infinite-dimensional symmetry that is generated by a non-Abelian loop algebra. This means that, as it happens in Einstein--Yang-Mills theory in asymptotically flat spacetimes \cite{GlennPH}, a Virasoro--Kac-Moody structure emerges in the near horizon region of colored black holes. In other words, the same symmetry enhancement phenomenon of the non-Abelian algebra discovered by Barnich and Lambert at null infinity also occurs near the black hole event horizon. Actually, the full algebra we will encounter in the near horizon limit differs from the one found near null infinity, the difference being the structure constants that connect supertranslations to superrotations, cf. \cite{DGGP3}; however, the Virasoro--Kac-Moody piece matches exactly. 

The paper will be organized as follows: In Section 2, we will consider the near horizon symmetries in Einstein--Yang-Mills theory. We will present a sensible set of boundary conditions at the horizon that, on the one hand, permit to accomodate the physically relevant solutions such as colored black holes and, on the other hand, turn out to be preserved by an infinite set of diffeomorphisms and gauge transformations. These boundary conditions are the generalization of those proposed in \cite{DGGP} to the non-Abelian case. In section 3, we will study the algebra of diffeomorphisms and gauge transformations preserving the prescribed boundary conditions, and we will show that it turns out to be an infinite-dimensional algebra that contains a Kac-Moody subalgebra. Section 4 contains a brief discussion about the relevance of our result.

\section{Non-Abelian horizons}

Let us consider a four-dimensional spacetime $(\mathcal{M}, g)$ with metric $ ds^2  =  g_{\mu \nu} \, dx^{\mu} dx^{\nu} $ (with $\mu , \nu = 0, 1, 2, 3$) and let us assume the spacetime has a non-singular, isolated, compact horizon $\mathcal{H} = \mathcal{H}^+\cup \mathcal{H}^-$ with $\mathcal{H}^{\pm }=\Sigma_2\times \mathbb{R}$, with $\Sigma_2$ being a compact spacelike 2-surface, say of topology $S^2$. We will consider advanced coordinates $x^0=v, \, x^A= z^A$ (with $A=1,2$) and $x^3=\rho$; the horizon is the hypersurface $\rho =0$ on which $v$ is null; $\Sigma_2$ is a constant-$v$ section of that hypersurface.

Without loss of generality, we can always choose coordinates such that, close to the future (past) horizon $\mathcal{H}^{+}=\Sigma_2\times \mathbb{R}$ ($\mathcal{H}^{-}$), the metric takes the form \cite{Booth:2012xm, Moncrief:2018tib}
\begin{align}
    g_{v v}=&\ -2\kappa\, \rho + \mathcal{O}(\rho^2), \label{av1} \\
    g_{v A}=&\ g_{v A}^{(1)}(z^B)\, \rho+\mathcal{O}(\rho^2), \label{av1_B}\\
    g_{AB}=&\ g_{AB}^{(0)}(z^C)+g_{AB}^{(1)}(z^C) \rho +\mathcal{O}(\rho^2),\label{av1_C}
\end{align}
together with the gauge fixing conditions
\begin{equation}\label{gauge}
    g_{\rho\rho}=0,\ \ \ g_{v \rho}=1,\ \ \ g_{A\rho}=0.
\end{equation}
The radial coordinate $\rho  \in \mathbb{R}_{\geq 0}$ measures the distance from the horizon, $v \in \mathbb{R}$ is null on $\mathcal{H}^{+}$, and $z^A$ (with $A=1,2$) represent coordinates on $\Sigma_2$. In (\ref{av1}), $g_{\mu \nu }^{(n)}$ stand for functions of $z^A $, each of which can be thought of as the coefficient of the order $\mathcal{O}(\rho ^n)$ in the near horizon (i.e. small $\rho $) expansion of the metric components. Constant $\kappa = -\frac 12 g^{(1)}_{v v }$ corresponds to the surface gravity of the horizon. We denote $\theta_A(z^B)\equiv g_{v A}^{(1)}(z^B)$ and $\Omega_{AB}(z^C)\equiv g_{AB}^{(0)}(z^C)$. The fact that the latter functions are independent of $v$ is guaranteed by the isolated horizon condition. 

We are interested in Einstein gravity coupled to Yang-Mills theory, and therefore we have to introduce, in addition to the metric, the non-Abelian gauge field $A_{\mu}^a$, which defines the gauge connection 1-form
\begin{equation}
    A = A_{\mu}^{a}\, T_a \, dx^{\mu}\, ,
\end{equation}
with $T_a$ being the generators of a Lie algebra g (with $a=1, 2, ... ,\, \dim(\text{g})$), which satisfy the Lie product
\begin{equation}
    \big[ T_b, T_c \big] = i f_{bc}^a T_a
\end{equation}
with $f_{bc}^a$ being the structure constants. Let $G$ be a compact, semisimple Lie group generated by g, which enters in the definition of the gauge theory through the standard building blocks: we have the covariant derivative $D_{\mu} = \partial_{\mu} - i \alpha A_{\mu}^a T_a$ and the field strength $F^a=F_{\mu \nu}^a = \partial_{\mu} A_{\nu}^a - \partial_{\nu} A_{\mu}^a + \alpha f_{bc}^a A_{\mu}^b A_{\nu}^c$ that yields the g-valued curvature 2-form $F^a_{\mu \nu }\, dx^{\mu}\wedge dx^{\nu}$; $\alpha $ is the gauge coupling. The action of the theory reads
\begin{equation}
I= \frac{1}{16\pi G}   \int_{\mathcal{M}} d^4x \sqrt{|g|} \, R \, -  \,\frac 14  \int_{\mathcal{M}} d^4x \sqrt{|g|} \, \text{Tr} F^2\, .
\end{equation}

Close to the horizon, the gauge field satisfies the following asymptotic behavior
\begin{align}
    &A_{v}^a = A_{v}^{(0)\, a} +   A_{v}^{(1)\, a}(v,z^{A})\, \rho + \mathcal{O}(\rho^2), \label{av2}\\
    &A_{B}^a = A_{B}^{(0)\, a}(z^{A}) +  A_{B}^{(1)\, a}(v,z^{A}) \, \rho + \mathcal{O}(\rho^2),\label{av2_B}
\end{align}
where $A_{\mu}^{(n)\, a}$ stands for the coefficient of the order $\mathcal{O}(\rho ^n)$ in the small $\rho $ expansion. In addition, the gauge condition $A_{\rho}^a=0$ holds for all $\rho $. Notice that, while $A_{\mu}^{(n>0)\, a}$ are functions of $z^A$ and $v$, $A_{v}^{(0)\, a}$ is constant and $A_{B}^{(0)\, a}$ only depends on $z ^A$; cf. \cite{DGGP3}. This suffices to accommodate charged black hole solutions within the configuration space.

Now, having defined the near horizon boundary conditions for both the gravitational and the gauge fields, let us consider the symmetry transformations that preserve such conditions. More precisely, we ask for those diffeomorphisms and gauge transformations that preserve the asymptotic form of (\ref{av1})-(\ref{av1_C}) and (\ref{av2})-(\ref{av2_B}), allowing for variation of the specific functions $g_{\mu\nu}^{(n)}$, $A_{\mu}^{(n)\, a}$ but preserving the functional form of the small $\rho $ expansion together with the gauge fixing conditions. Let us call $\chi_{\mu}$ and $\epsilon^a$ the asymptotic Killing vectors and the gauge parameters that realize such symmetry transformations, respectively. Then, we have the changes $g_{\mu\nu}\to g_{\mu\nu}+\delta_{\chi} g_{\mu\nu} $, $A_{\mu}^a\to A_{\mu}^a+\delta_{(\chi,\epsilon)} A_{\mu}^a $, with
\begin{equation}
    \delta_{\chi} g_{\mu \nu} = (\mathcal{L}_{\chi} g)_{\mu \nu}, \qquad \delta_{(\chi,\epsilon)} A_{\mu}^a = (\mathcal{L}_{\chi} A)_{\mu}^a + \partial_{\mu} \epsilon^a - \alpha f_{bc}^a \epsilon^b A_{\mu}^c \, .
\end{equation}
As said, we require these changes to be such that they preserve the asymptotic conditions prescribed above. For instance, in order to respect the gauge fixing conditions, we have to demand
\begin{equation}
    (\mathcal{L}_{\chi} g)_{\rho \rho} = 0, \quad (\mathcal{L}_{\chi} g)_{v \rho} = 0, \quad (\mathcal{L}_{\chi} g)_{A \rho} = 0, \quad (\mathcal{L}_{\chi} A)_{\rho}^a + \partial_{\rho}\epsilon^a= 0,
\end{equation}
where we have used $A_{\rho}^a=0$. The latter conditions, together with some closure conditions for the asymptotic expansion, yield the following form for $\chi$ and $\epsilon^a$
\begin{align}
&\chi^{v} = f(v,z^A)\\
&\chi^{\rho} =  - \rho \partial_{v} f + \partial_{A} f \int_{0}^{\rho} g^{AB} g_{v B} d\rho'\\
&\chi^{A} = Y^{A}(z^A) - \partial_{B} f \int_{0}^{\rho} g^{AB} d\rho'
\end{align}
together with
\begin{align}
&\epsilon^a = \epsilon_0^a(v,z^A) - \int_{0}^{\rho} A_{B}^a \partial_{\rho} \chi^B d\rho'\, .
\end{align}
Here, $f(v,z^A)$ is a function of $z^A$ whose dependence with $v$ will be later restricted by requiring extra conditions. Being the $v$-component of the asymptotic Killing vector, this function will ultimately be associated to horizon supertranslations. Function $\epsilon_0^a(v,z^A)$ depends on $z^A$ and $v$, while $Y^A(z^B)$ are two arbitrary functions of $z^B$ only ($A,B=1,2$). On $\Sigma_2$ we can consider complex coordinates $(z^1, z^2)=(z,\bar{z})$ and holomorphic and anti-holomorphic fields obeying $\bar{\partial }Y^z(z)= \partial Y^{\bar{z}}(\bar{z})=0$; this allows us to denote ${Y}=Y^{{z}}({z})$ $\bar{Y}=Y^{\bar{z}}(\bar{z})$ for short. The latter functions will be identified as those generating local conformal transformations on $\Sigma_2$, i.e. they will be related to the so-called supertranslations. 

In order to satisfy the boundary conditions (\ref{av2})-(\ref{av2_B}), the gauge field has to obey
\begin{equation}
    (\mathcal{L}_{\chi} A)_{v}^a + \partial_{v}\epsilon^a - \alpha f_{bc}^a\, \epsilon^b A_{v}^c= \mathcal{O}(\rho), \qquad (\mathcal{L}_{\chi} A)_{B}^a + \partial_{B}\epsilon^a - \alpha f_{bc}^a\, \epsilon^b A_{B}^c= \mathcal{O}(1)\, .
\end{equation}
Satisfying the first of these conditions implies solving a set of differential equations for $\epsilon_0^a$; namely 
\begin{equation}\label{dif}
    \partial_{v}(\chi^{v} A_{v}^{(0) \,a} + \epsilon_0^a) - \alpha f_{bc}^a \, \epsilon_0^b A_{v}^{(0) \, c} = 0
\end{equation}
where the indices $v$ are not contracted.

By computing the variations of the metric and the gauge field under diffeomorphisms and gauge transformations defined by the $\chi $ and $\epsilon $ given above, one finds
\begin{align}
    \delta_{(\chi,\epsilon)} \kappa &= \kappa \partial_{v} f + \partial_{v}^2 f = 0\label{la15}\\
    \delta_{(\chi,\epsilon)} \theta_{A} &= \mathcal{L}_{Y} \theta_{A} + f \partial_{v} \theta_{A} - 2 \kappa \partial_{A} f - 2 \partial_{v} \partial_{A}f + \Omega^{BC} \partial_{v} \Omega_{AB} \partial_{C} f\\
     \delta_{(\chi,\epsilon)} \Omega_{AB} &= f \partial_{v} \Omega_{AB} + \mathcal{L}_{Y} \Omega_{AB}\\
     \delta_{(\chi,\epsilon)} A_{v}^{(0)a} &= 0\\
     \delta_{(\chi,\epsilon)} A_{B}^{(0)a} &= Y^{C} \partial_{C}A_{B}^{(0)a} + A_{C}^{(0)a} \partial_{B} Y^{C} + \partial_B f A_{v}^{(0)a} + \partial_{B} \epsilon_0^a - \alpha f_{bc}^a \epsilon_0^b A_{B}^{(0)c}
\end{align}
Notice that in (\ref{la15}) we are additionally demanding the variation $\delta_{(\chi,\epsilon)} \kappa$ to vanish; that is to say, we are considering a phase space defined by functional variations that preserve the surface gravity. For the case of non-extremal horizons ($\kappa \neq 0$) this yields $f (v,z^A) = T(z^A) + e^{-\kappa v} X(z^A)$, with $T(z^A)$ and $X(z^A)$ being two arbitrary functions on $\Sigma_2$; cf. \cite{DGGP}.

\section{Symmetry algebra}

Now, let us derive the algebra that generates the asymptotic symmetries defined above. In order to obtain this algebra, let us start by looking at how two subsequent variations act on the metric function $\theta_A$; namely
\begin{equation}
    \big[\delta_{(\chi_1,\epsilon_1)}, \delta_{(\chi_2,\epsilon_2)} \big] \theta_A =  \tilde{Y}^A \theta_A + \tilde{f} \partial_{v} \theta_A - 2 \kappa \partial_A \tilde{f} - 2 \partial_{v} \partial_{A} \tilde{f} + \Omega^{BC} \partial_{v} \Omega_{AB} \partial_{C} \tilde{f}\label{GHJGJHJ}
\end{equation}
where
\begin{equation}
    \tilde{Y}^A = Y_1^B \partial_B Y_2^A - Y_2^B \partial_B Y_1^A, \quad \tilde{f} = \mathcal{L}_{Y_1} f_2 - \mathcal{L}_{Y_2} f_1 + f_1 \partial_{v} f_2 - f_2 \partial_{v} f_1.
\end{equation}
The brakets $[\, , \, ]$ in (\ref{GHJGJHJ}) is the modified Lie brakets introduced in \cite{Cedric}, which is valid in the case the gauge parameters are field dependent; see Eq. (2.4) therein; see also \cite{GlennPH, DGGP2} and references thereof.

Now, in order to determine the expression for $\tilde{\epsilon_0^a}$, it is sufficient to consider the variation of $A_{B}^{(0)a}$ and then see how the algebra closes. This yields the cumbersome expression
\begin{align}
    &\big[\delta_{(\chi_1,\epsilon_1)}, \delta_{(\chi_2,\epsilon_2)} \big] A_{B}^{(0)a} = \tilde{Y}^C \partial_C A_{B}^{(0)a} + A_{C}^{(0)a} \partial_B \tilde{Y}^C + \partial_B \tilde{f} A_{v}^{(0)a} +\partial_B \big[ Y_1^C \partial_C\, \epsilon_{0_2}^a - Y_2^C \partial_C\, \epsilon_{0_1}^a \big]\nonumber \\ 
    &- \alpha f_{bc}^a \big[ Y_1^C \partial_C\, \epsilon_{0_2}^b - Y_2^C \partial_C\, \epsilon_{0_1}^b \big] A_{B}^{(0)c} - \partial_B(f_1 \partial_{v} f_2 - f_2 \partial_{v} f_1) A_{v}^{(0)a}\label{24} \\
    \nonumber
    &- \alpha f_{bc}^a \big[ \epsilon_{0_1}^b (\partial_B f_2 A_{v}^{(0)c} + \partial_B \epsilon_{0_2}^c - \alpha f_{jk}^c \epsilon_{0_2}^j A_{B}^{(0)k}) - \epsilon_{0_2}^b (\partial_B f_1 A_{v}^{(0)c} + \partial_B \epsilon_{0_1}^c - \alpha f_{jk}^c \epsilon_{0_1}^j A_{B}^{(0)k}) \big]
\end{align}
where we have identified some terms, and added and subtracted $\partial_B(f_1 \partial_{v} f_2 - f_2 \partial_{v} f_1) A_{v}^{(0)a}$ in order to complete the expression of $\tilde{f}$. 

Using equation (\ref{dif}) to write $\partial_{v} f A_{v}^{(0)a} = \alpha f_{bc}^a \, \epsilon_0^b A_{v}^{(0) \, c}  - \partial_{v} \epsilon_0^a$, the Jacobi identity to write $f^a_{bc}\big[ f^c_{mn}(\epsilon_{0_1}^b \epsilon_{0_2}^m - \epsilon_{0_1}^m \epsilon_{0_2}^b) A^{(0) n}_B \big] = f^a_{bc} f^b_{mn} \epsilon_{0_1}^m \epsilon_{0_2}^n A^{(0) c}_B$, and defining $E^a \equiv Y_1^C \partial_C\, \epsilon_{0_2}^a - Y_2^C \partial_C\, \epsilon_{0_1}^a + f_1 \partial_{v}\epsilon^a_{0_{2}} - f_2 \partial_{v}\epsilon^a_{0_{1}}$ for short, the expression above can be rewritten as follows
\begin{align}
    &\big[\delta_{(\chi_1,\epsilon_1)}, \delta_{(\chi_2,\epsilon_2)} \big] A_{B}^{(0)a} = \tilde{Y}^C \partial_C A_{B}^{(0)a} + A_{C}^{(0)a} \partial_B \tilde{Y}^C + \partial_B \tilde{f} A_{v}^{(0)a}\nonumber\\ 
    &+\partial_B (E^a - \alpha f^a_{bc} \epsilon_{0_1}^b \epsilon_{0_2}^c)
    - \alpha f_{bc}^a (E^b - \alpha f^b_{mn} \epsilon_{0_1}^m \epsilon_{0_2}^n ) A_{B}^{(0)c}\label{general}\\
    \nonumber
    &+ f_1 \partial_{v} [\partial_B(\epsilon_{0_2}^a  + f_2 A_{v}^{(0)a}) - \alpha f^a_{bc} \epsilon_{0_2}^b A_{B}^{(0)c} ] - f_2 \partial_{v} [\partial_B(\epsilon_{0_1}^a  + f_1 A_{v}^{(0)a}) - \alpha f^a_{bc} \epsilon_{0_1}^b A_{B}^{(0)c} ]\, .
\end{align}

We are interested in writing this expression as 
\begin{equation}
     \delta_{(\tilde{\chi},\tilde{\epsilon})} A_{B}^{(0)a} = \tilde{Y}^{C} \partial_{C}A_{B}^{(0)a} + A_{C}^{(0)a} \partial_{B} \tilde{Y}^{C} + \partial_B \tilde{f} A_{v}^{(0)a} + \partial_{B} \tilde{\epsilon_0^a} - \alpha f_{bc}^a \tilde{\epsilon_0^b} A_{B}^{(0)c}
\end{equation}
for a given $\tilde{\epsilon_0^b}$. Using that $A_{B}^{(0)a}$ and $ Y^A $ only depend on $z^A$, we can freely add the term $\partial_{v} [Y^{C} \partial_{C}A_{B}^{(0)a} + A_{C}^{(0)a} \partial_{B} Y^{C}]$ to (\ref{general}) and then identify the last line of that equation as $f_1 \partial_v (\delta_{(\tilde{\chi_2},\tilde{\epsilon_2})}A_{B}^{(0)a}) - f_2 \partial_v (\delta_{(\tilde{\chi_1},\tilde{\epsilon_1})}A_{B}^{(0)a})$, which vanishes as $\delta_{(\tilde{\chi},\tilde{\epsilon})}A_{B}^{(0)a}$ only depends on $z^A$. This permits to identify $\tilde{\epsilon_0^a} = E^a - \alpha f^a_{bc} \epsilon_{0_1}^b \epsilon_{0_2}^c$, which reduces to the expression of \cite{DGGP3} in the Abelian case $f^{ab}_{\, c}=0$. In the most general case, we find the following set of variations
\begin{align}
    &\tilde{Y}^A = Y_1^B \partial_B Y_2^A - Y_2^B \partial_B Y_1^A \label{algebra0}\\
    &\tilde{f} = Y_1^C \partial_C f_2 - Y_2^C \partial_C f_1 + f_1 \partial_{v} f_2 - f_2 \partial_{v} f_1 \label{algebra0_B}\\
    &\tilde{\epsilon_0^a} = Y_1^C \partial_C\, \epsilon_{0_2}^a - Y_2^C \partial_C\, \epsilon_{0_1}^a + f_1 \partial_{v}\epsilon^a_{0_{2}} - f_2 \partial_{v}\epsilon^a_{0_{1}} - \alpha f^a_{bc} \epsilon_{0_1}^b \epsilon_{0_2}^c\label{algebra0_C}
\end{align}

Now, we can solve equation (\ref{dif}) and replace its solution in (\ref{algebra0_C}). Equation (\ref{dif}) can be written as $\partial_{v}(\epsilon^a_0 + f A^{(0) a}_{v}) = \Delta^a_b \epsilon^b_0 $ with $ \Delta^a_b = \alpha f^a_{bc} A^{(0)c}_{v}$, which has solution of the form
\begin{equation}
    \epsilon^a_0 (v,z^A) = \big[ e^{\Delta v} \big]^a_b 
    U^b(z^A) - f(z^A,v) A^{(0) a}_{v}\, .\label{La322}
\end{equation}
This generalizes the result $\epsilon^a_0 (v,z^A) = U^a(z^A) - f(z^A,v) A^{(0) a}_{v}$ with $a=1, 2, ...N$ obtained in \cite{DGGP3} for $G=U(1)^N$. The novel feature in (\ref{La322}) relative to the Abelian case is the $v$-dependent exponential accompanying the function $U^a(z^A)$, which obviously vanishes when the structure constants vanish. 

Having fully determined the functional dependence with the advanced null coordinate $v$, it is possible to obtain the expression for $\tilde{U}^a$ explicitly; namely
\begin{align}
    &\tilde{\epsilon_0^a} + \tilde{f} A^{(0) a}_{v} = \big[ e^{\Delta v} \big]^a_j
    \tilde{U}^j(z^A) = \big[ e^{\Delta v} \big]^a_j \big[Y_1^C \partial_C\, U_{2}^j - Y_2^C \partial_C\, U_{1}^j \big]
    -\alpha f^a_{bc} \big[ e^{\Delta v} \big]^b_k \big[ e^{\Delta v} \big]^c_m U^k_1 U^m_2
\end{align}
where we see that the dependence of $f_1 $ and $ f_2$ has cancelled out. We can use that $
    f^a_{bc} \big[ e^{\Delta v} \big]^b_k \big[ e^{\Delta v} \big]^c_m = f^j_{km} \big[ e^{\Delta v} \big]^a_j $, which follows from standard formulae applied to the adjoint representation. Taking all this into account, one finally obtains
\begin{equation}\label{GGG}
    \tilde{U}^a = Y_1^C \partial_C\, U_{2}^a - Y_2^C \partial_C\, U_{1}^a - \alpha f^a_{bc} U^b_1 U^c_2
\end{equation}
which, together with the transformations
\begin{align}
    &\tilde{Y}^A = Y_1^B \partial_B Y_2^A - Y_2^B \partial_B Y_1^A\label{algebra1}\\
    &\tilde{T} = Y_1^B \partial_B T_2 - Y_2^B \partial_B T_1\label{algebra1_B}\\
    &\tilde{X} = Y_1^B \partial_B X_2 - \kappa T_1 X_2 - Y_2^B \partial_B X_1 + \kappa T_2 X_1\label{algebra1_C}
\end{align}
form the symmetry algebra. Functions $Y^B$, $T$, $X$, and $U^a$ are four arbitrary functions of the coordinates $z^A$ on $\Sigma_2$, and this gives rise to an infinite-dimensional algebra. While the line (\ref{algebra1}) expresses the existence of two copies of the Witt algebra, namely two Virasoro algebras with vanishing central term, the line (\ref{algebra1_B}) gives the semidirect sum of the Virasoro algebras with an Abelian infinite-dimensional algebra, i.e. the so-called supertranslations. The line (\ref{algebra1_C}) shows the presence of another infinite-dimensional Abelian component, which acts on supertranslations as dilations act on standard translations. The new result here is (\ref{GGG}). This expresses, on the one hand, the mixing between the spacetime superrotations and the gauge transformations at the horizon; on the other hand, it generalizes the result of \cite{DGGP3} to the non-Abelian case, where the $G$-structure appears in the last term with the structure constants.

All the arbitrary functions of $z^A$ can be expanded in Fourier modes, e.g. as usually done when representing $\text{Diff}(S^1)$ or tensored $C^{\infty }(S^1)$ algebras in conformal field theory: It amounts to choose complex coordinates $z^A = (z,\bar{z})$ with $z=e^{\tau +i\sigma }$, and evaluate (\ref{GGG})-(\ref{algebra1}) on arbitrary modes $z^m \bar{z}^n$. That is to say, we can define $T(z,\bar{z}) = \kappa \sum_{m,n} T_{(m,n)} z^m \bar{z}^n$, $X(z,\bar{z}) = \sum_{m,n} X_{(m,n)} z^m \bar{z}^n$, $Y(z) = \sum_{n} Y_n z^n$, $\bar{Y}(\bar{z}) = \sum_{n} \bar{Y}_n \bar{z}^n$ and $U^a(z,\bar{z}) = \alpha  \sum_{m,n} U^a_{(m,n)} z^m \bar{z}^n$, and then express the algebra as follows
\begin{align}
    [Y_m, Y_n] &= (m-n) Y_{m+n}, \qquad [\bar{Y}_m, \bar{Y}_n] = (m-n) \bar{Y}_{m+n},\label{HJKM}\\
    [Y_k, T_{(m,n)}] &= - m T_{(m+k,n)}, \qquad [\bar{Y}_k, T_{(m,n)}] = - n T_{(m,n+k)},\\
    [Y_k, X_{(m,n)}] &= - m X_{(m+k,n)}, \qquad [\bar{Y}_k, X_{(m,n)}] = - n X_{(m,n+k)},\\
    [Y_k, U^a_{(m,n)}] &= - m U^a_{(m+k,n)}, \qquad [\bar{Y}_k, U^a_{(m,n)}] = - n U^a_{(m,n+k)},\\
    [X_{(k,l)},T_{(m,n)}] &=  X_{(m+k,n+l)}, \qquad [U^a_{(k,l)}, U^b_{(m,n)}] =  f^{ab}_{\ \, c}\, U^{c}_{(k+m,l+n)} \, ;\label{KM}
\end{align}
the other commutators vanish.

The last Lie product in (\ref{KM}) is our main result. This manifestly shows that in the near horizon region of black holes in Einstein--Yang-Mills theory the symmetry algebra gets enhanced in a way that it includes a (double) infinite-dimensional loop algebra $\hat{\text{g}}= \text{g}\otimes C^{\infty }(S^1)$ generated by $U^a_{(m,n)}$. This algebra is in semidirect sum with two copies of Witt algebra, being a spin-1 current under local conformal transformations. This generalizes the result found in \cite{DGGP3} for the $U(1)^N$ theory, which here corresponds to $f^{a}_{\, bc}=0$. The rest of the algebra matches the one found in \cite{DGGP, DGGP2}; namely: in addition to the loop algebra, the algebra above contains an infinite-dimensional Abelian piece generated by $T_{(m,n)}$, which is the horizon supertranslations. There is also an infinite-dimensional Abelian ideal generated by $X_{(m,n)}$. However, as shown in \cite{DGGP, DGGP2}, in the case of non-extremal black holes the latter algebra generates transformations that are pure gauge as $X(z,\bar{z})$ does not enter in the Noether charges associated to the asymptotic diffeomorphisms generated by $\chi = f(v, z,\bar{z})\partial_v$. We intend to return to the problem of computing the Noether charges associated to the full algebra (\ref{HJKM})-(\ref{KM}) and carefully analyzing their integrability conditions in a future work. Preliminary results on this permits to say that the conserved charges associated to $Y_{m}$, $\bar{Y}_{m}$ and $U^a_{(m,n)}$ form a Virasoro--Kac-Moody system with vanishing central charge and vanishing Kac-Moody level.  

\section{Discussion}

The results obtained here might have interesting applications to study colored black holes \cite{Bizon:1990sr, Volkov:1998cc}, cf. \cite{Bartnik:1988am}. In fact, it is possible to speculate that large gauge transformations could describe physical processes that produce a splash of colors on the horizon, for instance by means of a mechanism similar to the one studied in \cite{HPS2, DGGP3}. As in there, one can imagine dynamical processes that connect the asymptotic past null-infinity $\mathcal{I}^-$ with the future horizon $\mathcal{H}^+$, using that now we have learnt that a similar Virasoro--Kac-Moody structure emerges in both regions \cite{GlennPH, Campiglia, Banerjee}.

Besides, the formulae above can easily be extended to higher dimensions \cite{D}, and so applied to study selfgravitating Yang monopoles in arbitrary (even) dimension $D$. Such monopoles have been constructed in \cite{Gibbons:2006wd} for the gauge group $SO(D-2)$. For $D\geq 6$, such solutions describe non-Abelian black holes that, when the cosmological constant is negative, can have applications within the context of AdS/CFT correspondence, cf. \cite{Gibbons:2006wd}. It is possible to show that the analysis done in the previous sections applies to those non-Abelian solutions, and that the zero-mode of the Noether charges associated to the horizon supertranslations correctly reproduces the entropy of the black holes, cf. \cite{Wald2}. This is quite interesting since the selfgravitating Yang monopole has infinite gravitational energy and, therefore, the charge computation needed to study, for example, its thermodynamics only makes sense from the horizon perspective.

Other interesting questions related to the computation presented here relate to its possible generalizations. For example, one could ask whether a more general set of asymptotic conditions exist yielding non-vanishing central extensions of the charge algebra. We could also ask whether other algebraic structures, such as $w_N$ algebras, or even $w_{\infty +1}$ algebras as the recently found at null infinity \cite{W}, can also be realized at the horizon provided one prescribes adequate asymptotic conditions. These are all interesting questions for future investigations.

\[\]

The authors are indebted to Mauricio Leston for discussions. This work is based on previous collaborations with Laura Donnay and Hern\'an Gonz\'alez; G.G. thanks them for many enjoyable and illuminating conversations. This work has been partially supported by CONICET and ANPCyT through grants PIP-1109-2017, PICT-2019-00303.

\end{document}